\documentclass[a4paper]{article}

\usepackage[letterpaper]{geometry}
\usepackage[parfill]{parskip}
\PassOptionsToPackage{numbers, compress}{natbib}
\usepackage[numbers, compress]{natbib}
\usepackage{xr-hyper,refcount}

\usepackage{graphicx}
\usepackage[utf8]{inputenc} 
\usepackage[T1]{fontenc}    
\usepackage{url}  
\usepackage{tabularx}
\usepackage{booktabs}       
\usepackage{amsfonts}       
\usepackage{nicefrac}       
\usepackage{microtype}      
\usepackage[parfill]{parskip}
\usepackage{algorithm,algorithmic}
\usepackage{amsmath,amsthm,amssymb,bbm}
\usepackage{mathtools}
\usepackage{cases}
\usepackage{comment}
\usepackage{subcaption}
\usepackage{algorithm,algorithmic}
\usepackage{color}
\usepackage{appendix}
\usepackage{xspace}
\usepackage{enumitem}
\usepackage[english]{babel}
\usepackage{authblk}

\usepackage[colorlinks,citecolor=blue,urlcolor=blue,linkcolor=blue,linktocpage=true]{hyperref}
\usepackage{cleveref}
\crefformat{equation}{(#2#1#3)}
\crefrangeformat{equation}{(#3#1#4) to~(#5#2#6)}
\crefname{equation}{}{}
\Crefname{equation}{}{}

\crefname{definition}{\textbf{definition}}{definitions}
\Crefname{definition}{Definition}{Definitions}
\crefname{assumption}{\textbf{assumption}}{assumptions}
\Crefname{assumption}{Assumption}{Assumptions}
\definecolor{maroon}{RGB}{192,80,77}

\usepackage[utf8]{inputenc} 
\usepackage[T1]{fontenc}    
\usepackage{hyperref}       
\usepackage{url}            
\usepackage{booktabs}       
\usepackage{amsfonts}       
\usepackage{nicefrac}       
\usepackage{microtype}      
\usepackage{tikz}
\usepackage{tabularx}
\usetikzlibrary{automata,decorations.markings,positioning,arrows}

\usepackage{natbib}
\usepackage{amsmath}
\usepackage{bbm}

\usepackage[normalem]{ulem}

\usepackage{mathtools}

\usepackage{amssymb,amsmath,amsthm}
\usepackage{soul}
\usepackage{url}            
\usepackage{xcolor}
\usepackage{xr-hyper,refcount}
\usepackage[colorlinks]{hyperref}  

\title{Resolving the Human Subjects Status of Machine Learning’s Crowdworkers}

\begin{document}

\author{
  Divyansh Kaushik$^{\ddagger}$, Zachary C. Lipton$^\dagger$, Alex John London$^\dagger$\\
  $^\dagger$Carnegie Mellon University, $^\ddagger$Federation of American Scientists\\
  \href{mailto:dkaushik@fas.org}{\nolinkurl{dkaushik@fas.org}}, \{\href{mailto:zlipton@cmu.edu}{\nolinkurl{zlipton}},
  \href{mailto:ajlondon@cmu.edu}{\nolinkurl{ajlondon}}\}@andrew.cmu.edu
}

\maketitle

\begin{abstract}
In recent years, machine learning (ML) 
has relied heavily on crowdworkers
both for building datasets and 
for addressing research questions 
requiring human interaction or judgment. 
The diverse tasks performed and 
uses of the data produced
render it difficult to determine 
when crowdworkers are best thought of as workers (versus human subjects).
These difficulties are compounded by conflicting policies, 
with some institutions and researchers 
regarding all ML crowdworkers as human subjects
and others holding that they rarely constitute human subjects. 
Notably few ML papers involving crowdwork mention IRB oversight,
raising the prospect of non-compliance 
with ethical and regulatory requirements.
We investigate the appropriate designation of ML crowdsourcing studies, 
focusing our inquiry on natural language processing 
to expose unique challenges for research oversight.
Crucially, under the U.S. Common Rule, 
these judgments hinge on determinations of \emph{aboutness},
concerning both whom (or what) the collected data is about
and whom (or what) the analysis is about.
We highlight two challenges posed by ML:
the same set of workers can serve multiple roles 
and provide many sorts of information;
and ML research tends to embrace a dynamic workflow,
where research questions are seldom stated ex ante
and data sharing opens the door for future studies 
to aim questions at different targets. 
Our analysis exposes a potential loophole in the Common Rule, 
where researchers can elude research ethics oversight
by splitting data collection and analysis into distinct studies. 
Finally, we offer several policy recommendations 
to address these concerns.
\end{abstract}
\section{Introduction}
\label{sec:introduction}
As the focus of machine learning (ML)---and, 
in particular, natural language processing (NLP)---has 
shifted towards settings characterized by massive datasets,
researchers have become reliant on crowdsourcing platforms 
\citep{kovashka2016crowdsourcing,vaughan2017making,sheng2019machine,drutsa2021crowdsourcing}.
These practices have produced
hundreds of new datasets.
In NLP, for the task of passage-based question answering (QA) alone,
over $15$ new datasets containing at least $50k$ annotations
have been introduced since 2016.  
Prior to 2016, the available QA datasets 
contained at most an order of magnitude 
fewer human-annotated examples. 
The ability to construct such enormous resources derives, 
in large part, from the liquid market for temporary labor
enabled by crowdsourcing platforms, including 
Amazon Mechanical Turk, Upwork, Appen, and Prolific.
Over time, the relationship between the ML community 
and crowdworkers has evolved 
to encompass
a wide variety of tasks and interaction mechanisms.
However, the positive view of crowdsourcing 
as a means to produce \emph{better} and \emph{larger} datasets,
potentially leading to technological breakthroughs,
has been offset by growing concerns 
about the ethical and social dimensions 
of these one-off engagements with crowdworkers.
Points of concern include 
(i) the low wages received by crowdworkers 
\citep{fort2011last, whiting2019fair, silberman2018responsible, kummerfeld2021quantifying};
(ii) disparate access, benefits, 
and harms of developed applications
\citep{adelani2021masakhaner, nekoto2020participatory, orife2020masakhane, bender2018data, kiritchenko2018examining, rudinger2018gender, bender2021dangers, strubell2020energy};
(iii) the reproducibility of proposed methods 
\citep{dodge2019show, ning2020easy, freitag2021experts, card2020little};
and (iv) concerns about fairness and discrimination
arising in the resulting technologies
\citep{hovy2016social,leidner2017ethical, bender2020integrating, blodgett2020language}.

Our focus here is on what ethical framework should govern 
the interaction of ML researchers and crowdworkers, 
and the unique challenges posed by ML research to regulators.
While researchers in fields like NLP typically lack 
expertise in human subjects research,
they nevertheless require practical guidance 
for how to classify the role 
played by crowdworkers in their research 
so that they can comply with relevant
ethical and oversight requirements.
Unfortunately, clear guidance is presently lacking.
Reflecting the current state of confusion,
some institutions and a recent paper by Shmueli et al.~\cite{shmueli2021beyond} 
suggest that all ML crowdwork constitutes human subjects research,
while other institutions suggest that ML crowdworkers
rarely constitute human subjects~\cite{panos2009irb}.

In this paper, we address the source of confusion,
arguing that difficulties in resolving
the appropriate designation of ML's crowdworkers
owe to several formidable challenges:
\paragraph{Novel relationships} 
The ethical framework that oversight boards use 
to identify human subjects---the U.S. Common Rule---was 
developed in the wake of abuses 
in biomedical and behavioral research. 
This framework was especially influenced 
by dynamics in biomedical research, 
including the need to distinguish 
clinical research from medical practice \cite{london2021common}.
Binning activities into these categories
facilitated the goal of ensuring 
that these distinct relationships 
were governed by the relevant set of norms---the 
norms of clinical medicine 
or the norms of medical research.  
Because the distinction between employees on a research team 
and study participants is less ambiguous in medical contexts, 
little attention has been paid to criteria 
for distinguishing \emph{research staff} from study participants.

\paragraph{Novel methods} 
Compared to biomedical or social sciences, 
where data are collected to answer questions 
that have been specified in advance, 
ML research often involves a more dynamic workflow 
in which data are collected in an open-ended fashion 
and research questions are articulated 
in light of data and its analysis. 
Additionally, while it is typical in biomedicine 
for teams that gather data to analyze it, 
or for researchers to analyze data 
that was first gathered for clinical purposes, 
in ML research there can be a more distributed division of labor 
with some research groups collecting data 
that will serve as the foundation 
for future studies
by
a whole community of researchers.
\paragraph{Ambiguity}
Under the Common Rule,
whether an individual is a human subject hinges 
on whether the data collected, and later analyzed, 
is \emph{about} that individual.  
However, as Shmueli et al.~\cite{shmueli2021beyond} have noted, crowdworkers 
can fill such diverse roles in ML research (even within a single study)
that is becomes difficult to draw a line between 
which data is collected \emph{about} the crowdworker
versus merely \emph{from} them (but about something else) \cite{shmueli2021beyond}.
\paragraph{Inexperience} 
Despite the enormous productivity in this area,
crowdsourcing-intensive NLP papers 
seldom discuss the ethical considerations 
that would otherwise be central to human subjects research 
and rarely discuss whether an Institutional Review Board (IRB) 
approval or exemption was sought prior to the study---only 
$14$  ($\approx 2\%$) of the aforementioned $703$ papers
described IRB review or exemption \cite{shmueli2021beyond}\footnote{It 
is worth noting that in other computing fields 
such as human computer interaction, 
it is common practice to seek IRB review 
prior to collecting data from human annotators.};
\paragraph{Scale} Currently NLP research
is producing hundreds of crowdsourcing papers per year,
with 703 appearing at the top venues (ACL, NAACL, EMNLP) alone
from 2015--2020 \cite{shmueli2021beyond}.

Moreover, we argue that these challenges 
not only create confusion among stakeholders, 
they also open the potential for loopholes,
whereby researchers can avoid IRB oversight
without altering the substantive research procedures 
performed on participants \cite{london2020loopholes}.
In particular, a single study that would be considered 
human subjects research could be split into two parts: 
one in which researchers collect data about workers 
and release an anonymized version to the public 
without analyzing information about the workers themselves;
and a second in which they or another team of researchers 
perform analysis on information about the workers. 
According to some institutional policies, 
the latter two studies 
might not require research ethics oversight 
whereas the single study would.

To ensure that ML research is conducted 
according to the appropriate ethical and regulatory standards, 
greater clarity is required.  
In Section~\ref{sec:current_framework}, 
we elaborate the criteria that define human subjects 
for ethical and regulatory purposes in the United States.
We briefly discuss the relationship between 
the question of whether one or more persons satisfy these criteria 
and the question of whether that research must undergo review
by a properly constituted IRB.  
In Section~\ref{sec:examples}, we present 
prototypical examples from research in NLP 
to identify paradigmatic cases
for which it is clear/unclear 
how a given crowdworker should be classified.
We then show how the diversity of roles 
that crowdworkers can play in ML research 
poses a challenge for research ethics
and provide guidance on interpreting the Common Rule to identify 
when crowdworkers should be classified as human subjects 
versus as extensions of the research team 
for both ethical and regulatory purposes.
Finally, in Section \ref{sec:discussion}, 
we offer policy solutions 
to address these concerns.

\section{Current Regulatory Framework}
\label{sec:current_framework}
In the United States, the regulations that govern 
the use of human participants in scientific research 
are set out in the Code of Federal Regulations (CFR) 
and are commonly referred to as the Common Rule. 
These regulations are promulgated 
by the Executive Branch and
apply only to institutions that accept federal funds 
or that have agreed to abide by these rules. 
Nevertheless, the language and the requirements laid out in these rules 
have been adopted by, and exert a great deal of influence within, 
the larger literature on research ethics.  

Because the Common Rule only applies 
to research with human participants,
it sets out two important criteria 
to determine whether a person 
constitutes a research participant: 
those used to define \emph{research} 
and those used to define a \emph{human subject}.

First, in order to be a participant in research, 
research must be taking place.  
Research is defined, in part, 
as ``a systematic investigation, 
including research development,
testing, and evaluation, 
designed to develop or contribute
to generalizable knowledge.''
%
%
%
%
Second, human subjects 
are then defined as follows:

\emph{(e)(1) Human subject means 
a living individual about whom an investigator 
(whether professional or student) conducting research: 
\begin{enumerate}
    \item[(i)] Obtains information or biospecimens 
    through intervention or interaction with the individual, 
    and uses, studies, or analyzes 
    the information or biospecimens; or
    \item[(ii)] Obtains, uses, studies, analyzes, or generates 
    identifiable private information or identifiable biospecimens. (45 CFR 46.102 (e)(1))
\end{enumerate}}
\noindent For simplicity, we limit our discussion 
to the production of information, 
rather than to a discussion of specimens. 

Two points of clarification are in order. 
First, we note that in (e)(1), 
the definition of a human subject requires 
that researchers obtain information 
\emph{about} the individual in question.  
This does not imply that the researcher 
is conducting research about the individual, per se, 
since research aims to produce generalizable knowledge.
In the biomedical context, for example, 
a study might seek to determine the effect 
of some intervention on blood pressure 
among the population of individuals
who suffer from a particular disease.
To answer this question, researchers 
might 
measure the blood pressure
of specific individuals with that disease. 
That information is then analyzed 
to produce generalizable knowledge
pertaining to the underlying population.
However, as we will see, delineating precisely 
which information is \emph{about} an individual
can be difficult in many settings where crowdworkers
are engaged by ML researchers. 
Second, conditions (i) and (ii) lump together 
a range of cases that vary in substantive ways.
Condition (i) is a combination of two conjuncts.  
The first conjunct concerns the way that information is produced: 
information can arise from intervention or from interaction.  
These terms are defined respectively as:
\emph{
\begin{enumerate}
    \item[(2)] Intervention includes both physical procedures 
    by which information or biospecimens are gathered (e.g., venipuncture) 
    and manipulations of the subject or the subject’s environment 
    that are performed for research purposes.
    \item[(3)] Interaction includes communication or interpersonal contact 
    between investigator and subject.
\end{enumerate}
}
\begin{table*}[t!]
\centering
  \begin{tabularx}{15cm}{lXX}
    \toprule
& Studies/Analyzes & Uses \\
\midrule
Intervention & Identifying better crowdsourcing 
strategies via a randomized study & Train 
an ML model on data collected 
in a gamification environment \\
Interaction & Analyzing data collected via 
surveys on Mechanical Turk  & Asking crowd to annotate a dataset to train ML models \\
\bottomrule
  \end{tabularx}
  \caption{Examples of research interactions with the crowd.}
  \label{tab:interaction_type}
\end{table*}
Of these possibilities, interaction is the weaker condition.  
Interventions can reasonably be understood 
as the subset of interactions 
that produce a change in either the individual
(e.g., administering a drug, or drawing blood) 
or their environment 
(e.g., placing an individual in an imaging device). 
By contrast, interactions include communication 
or interpersonal contact 
that generate information 
without necessarily bringing about a 
change to the individual or their environment.  
For example, a study might involve 
randomizing a group of participants 
to receive either an investigational intervention 
in addition to usual care, or to receive only usual care.  
Although the former group receives 
an intervention---something they would 
not have received outside 
of the context of research---the latter group 
is not subject to an intervention.  
Nevertheless, their inclusion in a group 
that is randomized within a study constitutes 
a form of social interaction necessary 
to generate data that controls for confounding, 
and so helps to produce generalizable knowledge.

The second conjunct in condition (i) requires 
that information that arises in one of these two ways 
is then used, studied, or analyzed.  
Of these, \emph{use} is the broadest category, 
as there may be myriad ways information 
from a social interaction 
might be used in the course of research.  
In contrast, study and analysis 
seem to constitute a strict subset of uses 
in which data are analyzed or evaluated,
presumably to generate the generalizable knowledge 
that defines the study in question. 
Table \ref{tab:interaction_type} provides
a representation of the combinations of views 
that result from combining these modes 
of interaction and types of use. 
Among these, the \emph{intervention analysis} condition
is the most narrow and captures a paradigm 
of the researcher-participant relationship.  
Namely, a person is a human subject if, 
in the course of research, 
they are the target of an intervention 
from which a researcher generates information 
that is then the subject of an analysis 
that is intended to produce the generalizable information
that is the focus of the research.
In contrast, the \emph{interaction use} criteria are weaker,
holding that a person is human subject  
if, in the course of research, 
researchers interact with them in a way 
that produces information that is used
to further the goals of research.   

Condition (ii) deals with cases in which researchers obtain, 
use, study, analyze or generate private information about a living individual. 
This condition is intended to cover cases 
in which researchers might not interact with living persons,
in the sense outlined in condition (i), 
but they nevertheless use or generate private information 
about a living individual in the course of their research. 
This condition therefore applies to research involving datasets 
that include private information about living individuals 
or to research that would generate that information
from datasets that might not include private information 
about living individuals taken on their own.  

These definitions play a key role in demarcating 
which set of ethical and regulatory requirements 
apply to an activity. 
A research activity that does not involve human subjects 
does not fall under the purview of the regulations 
governing research with human subjects. 
Consequently, if there are no human subjects in a study 
then the study does not need to be reviewed by an IRB.  
In contrast, if a researcher is carrying out 
research with human participants, 
then that researcher incurs 
certain moral and regulatory responsibilities.  
Among these regulatory responsibilities is the duty 
to present one's research for review by an IRB. 

This last claim might come as a surprise 
to some who read the Common Rule, 
since a significant portion of ML research, 
and NLP research in particular, 
is likely to be classified as \emph{exempt}.  
Per 46.104.(3)(i), research involving benign behavioral interventions 
in conjunction with the collection of information 
from an adult subject through verbal 
or written responses (including data entry) 
or audiovisual recording can qualify for \emph{exempt} status
if the subject prospectively agrees
to the intervention and information collection 
and at least one of the following criteria is met:
\emph{
\begin{enumerate}
    \item[(A)] The information obtained is recorded by the investigator 
    in such a manner that the identity of the human subjects 
    cannot readily be ascertained, directly 
    or through identifiers linked to the subjects;
    \item[(B)] Any disclosure of the human subjects’ 
    responses outside the research 
    would not reasonably place the subjects 
    at risk of criminal or civil liability 
    or be damaging to the subjects’ financial standing, 
    employability, educational advancement, or reputation.
\end{enumerate}
}

However, a researcher cannot unilaterally 
declare their research to be \emph{exempt} from IRB review. 
Rather, \emph{exempt} is a regulatory category 
whose status must be certified by an IRB (§46.109.(a)). 
This can seem paradoxical to some since, 
in order to qualify for an exemption, 
researchers must submit sufficient information about their research 
to the IRB so that the latter can determine 
that these, or other applicable criteria 
(as laid out in the Common Rule) are met.\footnote{This is a commonality in administrative rulemaking as well as judicial review.
After all, Courts get to determine whether something is in their jurisdiction 
but a plaintiff has to provide information to enable a court to make that determination.} 
Nevertheless, the work required to secure this certification 
is usually less than is required to submit a full protocol 
and the certification is usually granted 
in less time than it would take for an IRB 
to provide a review of that protocol 
involving the full IRB board. 
For present purposes, the main point is that 
if a researcher at an institution 
bound by the Common Rule 
carries out human subjects research 
without first having that research 
reviewed by the relevant IRB,
then that researcher would be in violation 
of that institution's regulatory obligations, 
even if that research would have qualified for an exemption.  

\section{Common Rule and ML Research}
\label{sec:examples}

Based on the preceding analysis,
we can now identify a large subset of ML research 
in which crowdworkers are clearly human subjects.  
These cases fit squarely into the paradigm of research,
familiar in biomedicine and social science,
where researchers
interact with crowdworkers to produce data
\emph{about} those individuals,
and then analyze that data 
to produce generalizable knowledge
about a population from which those individuals
are considered to be representative samples.

First, we consider studies where researchers
assign crowdworkers at random to interventions 
in order to produce data that can be analyzed 
to generate generalizable knowledge 
about best practices for utilizing crowdworkers.
Here, the crowdworkers are clearly human subjects.
They are the target of an intervention
that was designed for the specific purpose
of capturing data about them
(namely, their performance at some task),
that could then be analyzed qualitatively and statistically
to address the central hypotheses of the study.   
%

For instance, consider Khashabi et al.~\cite{khashabi2020more},
who engage crowdworkers to investigate which workflows 
result in higher quality question-answering datasets. 
They recruit one set of crowdworkers 
to write questions given a passage, 
while another group of crowdworkers are shown a passage 
along with a suggested question 
and are tasked with minimally editing 
this question to generate new questions.
In these settings the data is about
the workers themselves, as is the analysis.
Investigating adversarial setups for
generating question answering datasets,
Kaushik et al.~\cite{kaushik2021efficacy} 
conduct a large-scale controlled study 
focused on a question answering task. 
One set of workers is asked to write five questions
after reading a passage, highlighting the answers to each, 
and are awarded a base pay of $\$0.15$ per question. 
Another set of crowdworkers is shown the same passages
but asked to write questions that elicit incorrect predictions
from an ML model trained using a different dataset
to perform passage based question answering.
To incentivize workers to spend more time 
thinking about ways to fool this existing model,
workers are paid $\$0.15$ 
for each question that fools the model 
in addition to the base pay of $\$0.15$ per question.
The research team later analyzed this data
to identify the differences 
between the questions generated by both sets of workers
and derive insights about how each data creation setup
influences crowdworker behavior.
They also trained various machine learning models on these datasets 
and evaluated them on several other question answering datasets 
to establish which 
interaction mechanism produced better data
(as measured by performance of models 
trained on the respective datasets), 
producing generalizable knowledge 
to aid future data collection efforts.

Humans subjects research in NLP is not limited
to studies aimed at dataset quality.
Hayati et al.~\cite{hayati2020inspired} 
paired two crowdworkers in a conversational setting 
and asked one crowdworker to 
recommend a movie to the other.
They then study the resulting data to identify 
what recommendation and communication strategies 
lead to successful recommendations,
and use these insights to train automated dialog systems.
In another work, Perez et al.~\cite{perez2015experiments}
asked crowdworkers to each write seven truths 
and seven plausible lies on topics of their own choice.
The authors also collected demographic attributes 
(such as age and gender) for each crowdworker.
They then analyzed how attributes of deceptive behavior 
relate 
to gender and age.
They also train classifiers using this data 
to predict deception, gender, and age.
In these cases, the researchers interacted 
with crowdworkers to produce data about the crowdworkers 
that was then analyzed to answer research hypotheses,
creating generalizable knowledge.


\subsection{Cases Where the Human Subjects Designation is Problematic}
\label{sec:problematic_cases}
Unlike the above, many ML crowdsourcing studies
do not fit neatly within the paradigm of research 
that is common in biomedicine and the social sciences. 
For example, crowdworkers are often brought in,
not as objects of study, but to perform tasks 
that could have been---and sometimes are---performed 
by members of the research team themselves.  
Note that in these cases, members of the research team
certainly do interact with crowdworkers 
and that those interactions produce data 
that in some meaningful sense is used to produce generalizable knowledge.
Moreover some of the collected data certainly
is \emph{about} the worker e.g., 
for the purpose of facilitating payment. 
However, in these cases, the data that is analyzed 
in order to produce generalizable knowledge 
are not  about the crowdworkers in any meaningful sense.

In perhaps the most common category of 
crowdsourcing study in machine learning,
researchers hire workers to label 
a training dataset that will be used
for training ML models.
For instance, 
Hovy et al.~\cite{hovy2014experiments} recruit crowdworkers
to annotate parts of speech in text.
They then train machine learning models on this data to predict parts of speech on test set.
In another study, Taboda et al.~\cite{taboada2011lexicon} 
recruit crowdworkers to create a 
collection of words associated with a sentiment label
which is then used to produce 
a sentiment classification model.
Countless such datasets are introduced every year.
Often researchers interact with the crowdworkers
and use the data generated as a result of that interaction. 
While it might appear that any such research satisfies
the \emph{interaction $+$ use} criteria from the Common Rule,
the subtle distinction is that 
the information used to produce generalizable knowledge
is not \emph{about} the worker. 
%
%

In many of these cases,
crowdworkers are performing tasks 
that are routinely performed
by research team members themselves
when working data on smaller scales. 
For example, Kovashka et al.~\cite{kovashka2016crowdsourcing}
describe numerous computer vision papers
where researchers provide their own labels.
In another example, NLP researchers 
often ask crowdworkers to not only 
provide the correct label for a document,
but also to highlight \emph{rationales},
contiguous segments in the text 
that provide supporting evidence.
Notably, while DeYoung et al.~\cite{deyoung2020eraser}
recruit crowdworkers to annotate rationales 
for various classification tasks,
Zaidan et al.~\cite{zaidan2007using}
opt to annotate the rationales themselves.
In another setting, Kaushik et al.~\cite{Kaushik2020Learning} recruit crowdworkers, 
who given a text and associated label, 
were tasked to minimally edit 
the text to make a counterfactual label applicable. 
In a followup study, instead of recruiting crowdworkers, 
Gardner et al.~\cite{gardner2020evaluating} 
opt to make these edits themselves.
%
%
%
%

How should crowdworkers in these cases be classified? 
On a strict reading of the claim that a human subject 
is a living individual ``about whom'' researchers obtain information
that is used or analyzed to produce generalizable knowledge,
then crowdworkers in these cases would not be classified as human subjects.  
This reading is consistent with the practice of some IRBs. 
For example, Whittier College's IRB states:\footnote{Archived on February 14, 2022. \href{https://web.archive.org/web/20220214194123/https://www.whittier.edu/academics/researchethics/irb/need}{https://web.archive.org/web/20220214194123/\\https://www.whittier.edu/academics/researchethics/irb/need}}
\begin{quote}
Information-gathering interviews with questions 
that focus on things, products, or policies 
rather than people or their thoughts about themselves 
may not meet the definition of human subjects research. 
Example: interviewing students about campus cafeteria menus 
or managers about travel reimbursement policy.
\end{quote}    

In contrast, other IRBs adopt a far more 
expansive reading of the language in the Common Rule. 
For instance, Loyola University's IRB says:\footnote{Archived on February 14, 2022. \href{https://web.archive.org/web/20220214194036/https://www.luc.edu/irb/irb_II.shtml}{https://web.archive.org/web/20220214194036/\\https://www.luc.edu/irb/irb\_II.shtml}}
\begin{quote}
In making a determination about 
whether an activity constitutes research involving human subjects, 
ask yourself the following questions:
\\
\emph{1) Will the data collected be publicly presented or published?}
\\
AND
\\
\emph{2) Do my research methods involve a) direct and/or indirect interaction 
with participants via interviews, assessments, surveys, or observations, 
or b) access to identifiable private information about individuals, e.g.,
information that is not in the public domain?}
\\
If the answer to both these questions is ``yes'', 
a project is considered research with human subjects 
and is subject to federal regulations.''
\end{quote}
\noindent Note that this interpretation
does not 
distinguish whether the information is about an individual 
or just obtained via a direct and/or indirect interaction. 
This view appears to be shared by other IRBs as well.\footnote{Archived on February 27, 2022. \href{https://web.archive.org/web/20220228012326/https://www.bsc.edu/academics/irb/documents/BSC\%20IRB\%20Decision\%20Tree.pdf}{https://web.archive.org/web/20220228012326/\\https://www.bsc.edu/academics/irb/documents/BSC\%20IRB\%20Decision\%20Tree.pdf}} 

\paragraph{How does information about versus merely from impact human subjects determination?} 
Traditionally, research ethics has not had to worry 
about who is a member of the research team 
and who is a participant in that research.  
This ambiguity arises in cases of self-experimentation,
but such cases are relatively rare
and fit squarely into the \emph{intervention $+$ analysis}
category from the Common Rule.  
The scope of the effort required to produce data 
that can be used in ML research 
has engendered new forms of interaction 
between researchers and the public.  
Without explicit guidance from federal authorities 
in the Office of Human Research Protections, 
individual IRBs will have to grapple with this issue on their own.

Our contention is that in the problematic cases
referred to in this section,
crowdworkers are best understood 
as augmenting the labor capacity of researchers 
rather than participating 
as human subjects in that research. 
This argument has two parts.

The first part is an argument from symmetry. 
Within a division of labor, if more than one person 
can carry out a portion of that division of labor, 
then the way that we categorize the activity in question 
should depend on substantive features of that activity 
rather than on the identity of the individual in question.\footnote{One might argue that the way we treat unionized vs non-unionized workers or independent contractors vs employees are counterexamples where the work might be exactly the same but the identity of the individual and a feature about them makes a difference regarding workplace protections amongst other things. In these cases, although, prior agreements might shape the entitlements of agents, they do not alter the classification of the activity performed i.e., whether the task is work or research.}  
From this, it follows that if a task is performed by a researcher 
in one instance and then by one or more crowdworkers in a second instance, 
then our categorization of that activity should be the same in both cases.  
The argument from symmetry alone entails only 
that either the crowdworker and the researcher 
are both part of the research team or both human subjects. 

The second part of the argument appeals 
to three additional considerations 
that support classifying both parties 
as part of the research team.  
First, when researchers perform the tasks in question 
it seems clear that they are not self-experimenting---they 
are not subjects in their own study. 
Second, this impression or intuition is explained 
by the fact that these interactions produce information 
that contributes to the production of generalizable knowledge, 
but that this information is better classified 
as coming from, rather than being about, these individuals.  
Researchers interact with other members of their team 
to produce information and this information is used in research,
but this use involves creating or refining 
the instruments, materials, metrics,
and other means necessary to carry out research. 
Its purpose is to create the means of generating new knowledge 
rather than to constitute that data or evidence base
whose study or analysis will generate this new knowledge.  
Third, ignoring the distinction between data 
that is about a person rather than merely from them,
and holding that both researchers and crowdworkers
are human subjects in these cases, 
creates a regulatory category so broad
that it would class members of every research team, 
including those in traditional biomedical and social science, as as human subjects. 
The reason is simply that those researchers 
routinely interact with other members of their team
to create information that is used 
to produce generalizable knowledge.   
But this consequence is absurd.

\subsection{Loopholes in Research Oversight}
The analysis in the previous section illustrates 
one challenge that ML research poses for research ethics.  
Part of the ethical rationale for the oversight of research with human participants is that the interests of study participants can be put at risk when researchers interact with or intervene upon them for the purpose of generating generalizable knowledge.  These risks can derive from the nature of the interaction or intervention, or from the use that is made of the resulting information.   
A loophole in research oversight has been defined ``as the unilateral ability of
a researcher to avoid an oversight requirement without
altering the substantive research procedures performed
on participants'' \cite{london2020loopholes}.  Loopholes in research oversight are morally troubling, in part, because they violate a concern about equal treatment for like cases: if researchers interact with individuals for the purpose of generating data that is about those individuals and generalizable knowledge is produced from the study or analysis of that data, then the interests of those individuals should receive the same level of oversight regardless of how the labor in this process is organized.  However, two features of ML research make the Common Rule particularly prone to loopholes: the way that labor is divided between the collection and the analysis of data and the way that research questions often arise after data collection.  

\paragraph{Scenario 1}
It is clear that the Common Rule envisions several ways in which labor might be divided in research.  First, in traditional biomedical or social science research it is common for the same individuals who collect data to also analyze that data in the course of their research.  This division of labor is presupposed in 
45 CFR 46.102 (e)(1)(i) which says that when a researcher
conducting research ``[o]btains information or biospecimens through intervention or interaction with the individual, \emph{and} uses, studies, or analyzes the information or biospecimens'', that research activity would be categorized as human subjects research. In this scenario, ethics review covers two morally weighty aspects of this division of labor: whether researchers interact with participants in ways that respect their autonomy and safeguard their welfare and whether they use the information that they obtain from these interactions in a way that respects the rights and welfare of the people this information is about. 

Second, it is common for data or biospecimens to be generated in the course of the provision of medical care or other health services.  In these cases the interactions of medical professionals with patients are not shaped around research purposes---they are shaped by the goals and purposes of the provision of health care or other medical services.  As such, those interactions are usually governed by the norms of medical or professional ethics.  Research ethics review thus focuses on whether the data or specimens in question constitute or include identifiable private information and, if so, whether research with this information respects the rights and welfare of the individuals from whom the information was gathered.  

It is not clear that the Common Rule envisioned a division of labor in which researchers would interact with individuals for research purposes (i.e., where the interactions are shaped by the goal of generating generalizable knowledge rather than the provision of health services) but those researchers would not use, study or analyze that information themselves.  To be clear, this is different from secondary use of data that was gathered for research purposes since, in traditional biomedical or behavioral research, the initial research would likely have been subject to research oversight. That oversight would ensure that researchers interact with participants on terms that respect their rights and welfare and subsequent oversight would evaluate additional uses of that data.  

In contrast, it is common for ML researchers to collect large datasets in an open-ended manner 
before hypotheses are formulated, 
often with the goal of facilitating 
a range of future research in broad topic areas
\cite{williams2018broad,zhang2021multimet, aggarwal2021explanations, ao2021pens, le2021dvd, zang2021photochat}.
For example, 
Williams et al.~\cite{williams2018broad} collect a large scale corpus
for the task of recognizing textual entailment.
They train an ML model on this dataset 
and release the dataset with anonymized crowdworker identifiers
for future research.
Similarly, Mihaylov et al.~\cite{mihaylov2018can} 
and Talmor et al.~\citep{talmor2019commonsenseqa} collect 
large scale question answering datasets created by crowdworkers,
train ML models on this data,
and release these datasets with anonymized crowdworker identifiers
for future research.
Since these studies only involved interacting with crowdworkers,
and using or analyzing data \emph{from} crowdworkers,
they may not require IRB review.
Subsequently, Geva et al.~\cite{geva2019we} took these anonymized datasets
and analyzed information \emph{about} the crowdworkers.
Specifically, they looked at how ML models trained on 
data created by one set
of crowdworkers do not generalize to 
data created by a disjoint set of crowdworkers.
They further train ML models, which given a document as input,
predict which crowdworker wrote that document.
Since Geva et al.~\cite{geva2019we} did not interact with the crowdworkers,
and only analyzed existing (anonymized) datasets, 
their studies also may not require IRB review.
However, 
had the researchers who collected these datasets also
analyzed this information, 
that study would have required IRB review.
As part of this review, an IRB would not only perform oversight over
the questions asked, but also how the researchers interact with the crowd
and whether adequate protections are in place for 
crowdworkers participating in these studies.

Although a significant portion of ML research poses few risks to participants, there are cases where interactions or interventions
are less benign, as when researchers ask crowdworkers to write toxic comments.
For example,
Xu et al.~\cite{xu2020recipes} recruit crowdworkers
to interact with an automated chatbot with the aim of eliciting 
\emph{unsafe} responses from the chatbot, 
using this data to train models that
are better at generating \emph{safe} responses. 
Crowdworkers may not be human subjects in this case, 
insofar as the information they provide 
is not about them in the relevant sense. 
However, in this example, the research team 
also created a taxonomy of offensive language types 
to classify human utterances citing potential use 
for this taxonomy in future research. 
From this larger data set inferences could be drawn 
about the proclivities to, or proficiency of, particular crowdworkers 
using offensive language of particular types.

In each of these cases, datasets are collected which contain information that is from crowdworkers for the purposes of producing generalizable knowledge that can include information that is about the crowdworkers.  A loophole in research oversight is created because   
45 CFR 46.102 (e)(1)(i) holds that 
individuals participating in a study are considered human subjects if
researchers both obtain \emph{and} use, study or analyze that information in a single study. To be clear, releasing such a dataset 
with identifiable private information 
for research purposes would fall under clause
(ii) from 45 CFR 46.102(e)(1) 
(discussed in Section \ref{sec:current_framework}). 
Once the dataset has been created, 
then using it for research purposes 
would fall under this same clause,
as long as the identifiable private information remains.  

A division of labor in which one set of researchers interact with individuals specifically for the purpose of generating data necessary to produce generalizable knowledge and then release that data (with anonymized identifiers) so that another set of researchers can  analyze it represents a loophole because, unlike the secondary use of data from ordinary clinical practice, this data is produced by researchers who interact with individuals for research purposes--to produce data that will help to create generalizable knowledge.  But, unlike the case where the researchers themselves analyze this data, this research activity would not be subject to oversight or review aimed at providing credible social assurance that those interactions respect individual autonomy and welfare \citep{london2021common}.   Anonymizing the data that is produced helps to shield individuals from harm that results from the way that information is used, such as uses that expose sensitive personal information.  But whether the means used to gather that data respect the autonomy and wellbeing of those individuals is not subject to oversight or review.

As a result, one way to address loopholes of this type would be to amend 45 CFR 46.102 (e)(1)(i) to explicitly include the \textbf{release} of data alongside its use, study or analysis.  

\paragraph{Scenario 2}
Amending 45 CFR 46.102 (e)(1)(i) to include the release of data may not be sufficient to foreclose a second scenario in which loopholes might arise.  Consider a scenario in which a research team interacts  
with crowdworkers to collect some data from and some that is about them and then proceeds to analyze both sets of data.  This single protocol fits the mold of traditional research in biomedicine or the social sciences and so would constitute research with human subjects.  Now consider a scenario in which the research team distributes this work over two separate protocols.
In the first protocol they propose to gather data 
that is both from and about crowd workers 
but only use data that is from them in their analyses. 
This study might not require IRB approval 
because it does not analyze, study or use data 
that is \emph{about} crowdworkers. 
The researchers then anonymize the full dataset
and submit a second protocol in which they analyze 
the now-anonymized data to answer questions about the crowd workers. 
The second study might not require IRB approval
because it does not involve obtaining information via any interaction with living individuals 
and it does not involve generating or using any identifiable private information.  

In this scenario, a single study that would require IRB approval 
could be decomposed into separate studies 
that involve the same interactions or interventions on crowdworkers in order to answer the same set of hypotheses 
but in a way that avoids research ethics oversight.  Because the researchers are not releasing their data publicly, the proposal in the previous section would not close this loophole.  
As a result, the determination of whether an ML project constitutes research with human participants might need to be made at a higher level than the individual study protocol. In the context of drug development, a trial \emph{portfolio} has been defined as a ``series of trials that are interrelated by a common set of objectives'' \cite{london2019clinical}.    
In ML research, the determination of whether an activity constitutes research with human participants may need to be made at the portfolio level by considering whether data to be generated and the questions to be investigated across an interrelated set of investigations are \emph{about} the crowdworkers. 
For portfolio level reviews to succeed, however, researchers would need to identify ex ante the scope and nature of the data they are collecting and the full range of questions they might seek to answer from that data across multiple studies.  Given the dynamic nature of ML research and the extent to which research questions are often posed after data has been collected, this may require consultation with IRBs to determine the conditions under which an envisioned portfolio of studies would or would not constitute research with humans and the steps that can be taken ex ante to facilitate the ability of researchers to pursue important questions as they arise.

\section{Discussion}
\label{sec:discussion}
There is currently considerable confusion about 
when ML's crowdworkers constitute human subjects 
for ethical and regulatory purposes.
While some sources suggest treating all crowdworkers as human subjects \citep{shmueli2021beyond},
our analysis makes a more nuanced proposal,
identifying:
(i) clear-cut cases of human subjects research: 
these require IRB consultation, 
even if only to confirm that they belong to an exemption category; 
(ii) crowdsourcing studies that do not constitute human subjects research 
because the analyses that produce generalizable information
do not involve data \emph{about} the workers; 
(iii) difficult cases, 
where the distinctive features 
of ML's crowdworking studies
combine with ambiguities in the Common Rule
to create substantial uncertainty
about how to apply existing requirements;
and (iv) loopholes, whereby researchers can escape
the human subjects designation without making
substantive changes to the research performed.

Part of the spirit of research oversight is to safeguard 
the rights and interests of individuals involved in research.  
In some cases, crowdworkers are the subjects of interventions or interactions that are designed to generate information about them which researchers intend to analyze in order to create generalizable knowledge.  
In these cases, the task of securing their rights and interests rightfully falls into the domain of human subjects ethics and oversight. 
But if researchers don't seek to either obtain 
or use, study, analyze or release information
\emph{about} a person (in some meaningful sense),
then it is not clear that frameworks for the protection of participants in research with human subjects are applicable or appropriate. 
Individuals who are not research participants 
can still be exposed to risks to their well-being and threats to their autonomy.  
This is true of most social interactions. 
It is particularly true of employment interactions 
as employers often have access to sensitive, private, identifiable information 
(such as Social Security Number, travel records, and background check reports) about their employees.
But the solution to ensuring that crowdworkers have credible public assurance that their rights and interests are protected is not to expand 
the definition of human subjects to include all crowdworkers.  
Rather, this goal should be achieved by reducing uncertainty about when crowdworkers constitute human subjects, ensuring proper research oversight when they do, and ensuring that in all other cases, crowdworker rights and interests are safeguarded through ethical and regulatory frameworks that govern employment relationships, workplace safety, and other labor practices.

\paragraph{Recommendations:}
\begin{enumerate}
    \item \textbf{ML researchers} must work proactively with IRBs 
    to determine which, if any, information they will generate 
    that is \emph{about} versus merely \emph{from} crowdworkers 
    and whether, given the full range of questions they intend 
    to investigate across the portfolio of studies involving this data, 
    the anticipated set of studies constitutes human subject research. 
    They should also recognize that 
    as the questions they investigate change, 
    the status of the research they are conducting 
    may change correspondingly.  
    Researchers should therefore work proactively 
    with their IRB to determine when modifications 
    to ongoing research require a new submission 
    or the submission of a protocol modification for IRB review.  
    \item \textbf{IRBs} should not reflexively
    classify all ML research involving crowdworkers as human subjects research.  
    At the same time, IRBs should also establish clear procedures 
    for evaluating portfolios of research to address 
    the possibility of loopholes in research oversight. 
    They should also communicate with ML researchers clearly
    about the conditions under which the classification of research might change
    and the conditions under which a revised protocol would need to be submitted.
    \item \textbf{The Office of Human Research Protections (OHRP)} 
    should offer more precise guidance about 
    what it means for information or analysis
    to be ``about'' a set of individuals. 
    We also recommend that OHRP should revise the Common Rule 
    so that 45 CFR 46.102(e)(1) condition (i) reads:
    ``Obtains information or biospecimens through intervention or interaction with the individual, and uses, studies, analyzes, \textbf{or releases} the information or biospecimens.''
    In short, this modification would require that an original investigator who collects data through interaction with humans and plans to release a dataset (even if anonymized) that could be used to ask questions \emph{about} those individuals
    must secure IRB approval for the research in which those data are gathered.
    Subsequent studies that draw upon the resulting anonymized public resource would not be marked as human subjects research, provided that they do not attempt to re-identify the individuals represented in the dataset. 
    %
    This modification would resolve the loophole identified in this paper.
    OHRP also has a role to play in offering guidance to ML researchers,
    which could be achieved by issuing 
    an agency Dear Colleague letter or an FAQ document. 
\end{enumerate}

\section*{Acknowledgements}
\label{sec:acknowledgements}
The authors thank Sina Fazelpour, Holly Fernandez Lynch and I Glenn Cohen for their constructive feedback. They also thank the CMU Block Center, the CMU PwC Center, Meta Research, and Amazon Research for the grants and fellowships that made this work possible.


\bibliographystyle{apa-good}
\bibliography{refs}

\begin{thebibliography}{49}
\expandafter\ifx\csname natexlab\endcsname\relax\def\natexlab#1{#1}\fi
\expandafter\ifx\csname url\endcsname\relax
  \def\url#1{{\tt #1}}\fi
\expandafter\ifx\csname urlprefix\endcsname\relax\def\urlprefix{URL }\fi

\bibitem[{Adelani et~al.(2021)Adelani, Abbott, Neubig, D'souza, Kreutzer,
  Lignos, Palen-Michel, Buzaaba, Rijhwani, Ruder
  et~al.}]{adelani2021masakhaner}
Adelani, D.~I., Abbott, J., Neubig, G., D'souza, D., Kreutzer, J., Lignos, C.,
  Palen-Michel, C., Buzaaba, H., Rijhwani, S., Ruder, S., et~al. (2021).
\newblock Masakhaner: Named entity recognition for african languages.
\newblock {\em arXiv preprint arXiv:2103.11811\/}.

\bibitem[{Aggarwal et~al.(2021)Aggarwal, Mandowara, Agrawal, Khandelwal,
  Singla, \& Garg}]{aggarwal2021explanations}
Aggarwal, S., Mandowara, D., Agrawal, V., Khandelwal, D., Singla, P., \& Garg,
  D. (2021).
\newblock Explanations for commonsenseqa: New dataset and models.
\newblock In {\em Workshop on Commonsense Reasoning and Knowledge Bases\/}.

\bibitem[{Ao et~al.(2021)Ao, Wang, Luo, Qiao, He, \& Xie}]{ao2021pens}
Ao, X., Wang, X., Luo, L., Qiao, Y., He, Q., \& Xie, X. (2021).
\newblock Pens: A dataset and generic framework for personalized news headline
  generation.
\newblock In {\em Proceedings of the 59th Annual Meeting of the Association for
  Computational Linguistics and the 11th International Joint Conference on
  Natural Language Processing (Volume 1: Long Papers)\/}, (pp. 82--92).

\bibitem[{Bender \& Friedman(2018)}]{bender2018data}
Bender, E.~M., \& Friedman, B. (2018).
\newblock Data statements for natural language processing: Toward mitigating
  system bias and enabling better science.
\newblock {\em Transactions of the Association for Computational
  Linguistics\/}, {\em 6\/}, 587--604.

\bibitem[{Bender et~al.(2021)Bender, Gebru, McMillan-Major, \&
  Shmitchell}]{bender2021dangers}
Bender, E.~M., Gebru, T., McMillan-Major, A., \& Shmitchell, S. (2021).
\newblock On the dangers of stochastic parrots: Can language models be too big?
\newblock In {\em Proceedings of the 2021 ACM Conference on Fairness,
  Accountability, and Transparency\/}, (pp. 610--623).

\bibitem[{Bender et~al.(2020)Bender, Hovy, \&
  Schofield}]{bender2020integrating}
Bender, E.~M., Hovy, D., \& Schofield, A. (2020).
\newblock Integrating ethics into the nlp curriculum.
\newblock In {\em Proceedings of the 58th Annual Meeting of the Association for
  Computational Linguistics: Tutorial Abstracts\/}, (pp. 6--9).

\bibitem[{Blodgett et~al.(2020)Blodgett, Barocas, Daum{\'e}~III, \&
  Wallach}]{blodgett2020language}
Blodgett, S.~L., Barocas, S., Daum{\'e}~III, H., \& Wallach, H. (2020).
\newblock Language (technology) is power: A critical survey of “bias” in
  nlp.
\newblock In {\em Proceedings of the 58th Annual Meeting of the Association for
  Computational Linguistics\/}, (pp. 5454--5476).

\bibitem[{Card et~al.(2020)Card, Henderson, Khandelwal, Jia, Mahowald, \&
  Jurafsky}]{card2020little}
Card, D., Henderson, P., Khandelwal, U., Jia, R., Mahowald, K., \& Jurafsky, D.
  (2020).
\newblock With little power comes great responsibility.
\newblock In {\em Proceedings of the 2020 Conference on Empirical Methods in
  Natural Language Processing (EMNLP)\/}, (pp. 9263--9274).

\bibitem[{DeYoung et~al.(2020)DeYoung, Jain, Rajani, Lehman, Xiong, Socher, \&
  Wallace}]{deyoung2020eraser}
DeYoung, J., Jain, S., Rajani, N.~F., Lehman, E., Xiong, C., Socher, R., \&
  Wallace, B.~C. (2020).
\newblock {ERASER: A Benchmark to Evaluate Rationalized NLP Models}.
\newblock In {\em Proceedings of the 58th Annual Meeting of the Association for
  Computational Linguistics\/}, (pp. 4443--4458).

\bibitem[{Dodge et~al.(2019)Dodge, Gururangan, Card, Schwartz, \&
  Smith}]{dodge2019show}
Dodge, J., Gururangan, S., Card, D., Schwartz, R., \& Smith, N.~A. (2019).
\newblock Show your work: Improved reporting of experimental results.
\newblock In {\em Proceedings of the 2019 Conference on Empirical Methods in
  Natural Language Processing and the 9th International Joint Conference on
  Natural Language Processing (EMNLP-IJCNLP)\/}, (pp. 2185--2194).

\bibitem[{Drutsa et~al.(2021)Drutsa, Ustalov, Fedorova, Megorskaya, \&
  Baidakova}]{drutsa2021crowdsourcing}
Drutsa, A., Ustalov, D., Fedorova, V., Megorskaya, O., \& Baidakova, D. (2021).
\newblock Crowdsourcing natural language data at scale: A hands-on tutorial.
\newblock In {\em Proceedings of the 2021 Conference of the North American
  Chapter of the Association for Computational Linguistics: Human Language
  Technologies: Tutorials\/}, (pp. 25--30).

\bibitem[{Fort et~al.(2011)Fort, Adda, \& Cohen}]{fort2011last}
Fort, K., Adda, G., \& Cohen, K.~B. (2011).
\newblock Last words: {A}mazon {M}echanical {T}urk: Gold mine or coal mine?
\newblock {\em Computational Linguistics\/}, {\em 37\/}(2), 413--420.
\newline\urlprefix\url{https://www.aclweb.org/anthology/J11-2010}

\bibitem[{Freitag et~al.(2021)Freitag, Foster, Grangier, Ratnakar, Tan, \&
  Macherey}]{freitag2021experts}
Freitag, M., Foster, G., Grangier, D., Ratnakar, V., Tan, Q., \& Macherey, W.
  (2021).
\newblock Experts, errors, and context: A large-scale study of human evaluation
  for machine translation.
\newblock {\em arXiv preprint arXiv:2104.14478\/}.

\bibitem[{Gardner et~al.(2020)Gardner, Artzi, Basmov, Berant, Bogin, Chen,
  Dasigi, Dua, Elazar, Gottumukkala et~al.}]{gardner2020evaluating}
Gardner, M., Artzi, Y., Basmov, V., Berant, J., Bogin, B., Chen, S., Dasigi,
  P., Dua, D., Elazar, Y., Gottumukkala, A., et~al. (2020).
\newblock Evaluating models’ local decision boundaries via contrast sets.
\newblock In {\em Proceedings of the 2020 Conference on Empirical Methods in
  Natural Language Processing: Findings\/}, (pp. 1307--1323).

\bibitem[{Geva et~al.(2019)Geva, Goldberg, \& Berant}]{geva2019we}
Geva, M., Goldberg, Y., \& Berant, J. (2019).
\newblock Are we modeling the task or the annotator? an investigation of
  annotator bias in natural language understanding datasets.
\newblock In {\em Proceedings of the 2019 Conference on Empirical Methods in
  Natural Language Processing and the 9th International Joint Conference on
  Natural Language Processing (EMNLP-IJCNLP)\/}, (pp. 1161--1166).

\bibitem[{Hayati et~al.(2020)Hayati, Kang, Zhu, Shi, \&
  Yu}]{hayati2020inspired}
Hayati, S.~A., Kang, D., Zhu, Q., Shi, W., \& Yu, Z. (2020).
\newblock Inspired: Toward sociable recommendation dialog systems.
\newblock In {\em Proceedings of the 2020 Conference on Empirical Methods in
  Natural Language Processing (EMNLP)\/}, (pp. 8142--8152).

\bibitem[{Hovy et~al.(2014)Hovy, Plank, \& S{\o}gaard}]{hovy2014experiments}
Hovy, D., Plank, B., \& S{\o}gaard, A. (2014).
\newblock Experiments with crowdsourced re-annotation of a pos tagging data
  set.
\newblock In {\em Proceedings of the 52nd Annual Meeting of the Association for
  Computational Linguistics (Volume 2: Short Papers)\/}, (pp. 377--382).

\bibitem[{Hovy \& Spruit(2016)}]{hovy2016social}
Hovy, D., \& Spruit, S.~L. (2016).
\newblock The social impact of natural language processing.
\newblock In {\em Proceedings of the 54th Annual Meeting of the Association for
  Computational Linguistics (Volume 2: Short Papers)\/}, (pp. 591--598).

\bibitem[{Ipeirotis()}]{panos2009irb}
Ipeirotis, P. (????).
\newblock {Mechanical Turk, Human Subjects, and IRB's}.
\newline\urlprefix\url{https://www.behind-the-enemy-lines.com/2009/01/mechanical-turk-human-subjects-and-irbs.html}

\bibitem[{Kaushik et~al.(2020)Kaushik, Hovy, \& Lipton}]{Kaushik2020Learning}
Kaushik, D., Hovy, E., \& Lipton, Z. (2020).
\newblock Learning the difference that makes a difference with
  counterfactually-augmented data.
\newblock In {\em International Conference on Learning Representations\/}.
\newline\urlprefix\url{https://openreview.net/forum?id=Sklgs0NFvr}

\bibitem[{Kaushik et~al.(2021)Kaushik, Kiela, Lipton, \&
  Yih}]{kaushik2021efficacy}
Kaushik, D., Kiela, D., Lipton, Z.~C., \& Yih, W.-t. (2021).
\newblock On the efficacy of adversarial data collection for question answering
  results from a large-scale randomized study.
\newblock In {\em Proceedings of the 59th Annual Meeting of the Association for
  Computational Linguistics and the 11th International Joint Conference on
  Natural Language Processing (ACL-IJCNLP)\/}.

\bibitem[{Khashabi et~al.(2020)Khashabi, Khot, \& Sabharwal}]{khashabi2020more}
Khashabi, D., Khot, T., \& Sabharwal, A. (2020).
\newblock More bang for your buck: Natural perturbation for robust question
  answering.
\newblock In {\em Proceedings of the 2020 Conference on Empirical Methods in
  Natural Language Processing (EMNLP)\/}, (pp. 163--170).

\bibitem[{Kiritchenko \& Mohammad(2018)}]{kiritchenko2018examining}
Kiritchenko, S., \& Mohammad, S. (2018).
\newblock Examining gender and race bias in two hundred sentiment analysis
  systems.
\newblock In {\em Proceedings of the Seventh Joint Conference on Lexical and
  Computational Semantics\/}, (pp. 43--53).

\bibitem[{Kovashka et~al.(2016)Kovashka, Russakovsky, Fei-Fei, \&
  Grauman}]{kovashka2016crowdsourcing}
Kovashka, A., Russakovsky, O., Fei-Fei, L., \& Grauman, K. (2016).
\newblock Crowdsourcing in computer vision.
\newblock {\em Foundations and Trends in Computer Graphics and Vision\/}, {\em
  10\/}(3), 177--243.

\bibitem[{Kummerfeld(2021)}]{kummerfeld2021quantifying}
Kummerfeld, J.~K. (2021).
\newblock Quantifying and avoiding unfair qualification labour in
  crowdsourcing.
\newblock {\em arXiv preprint arXiv:2105.12762\/}.

\bibitem[{Le et~al.(2021)Le, Sankar, Moon, Beirami, Geramifard, \&
  Kottur}]{le2021dvd}
Le, H., Sankar, C., Moon, S., Beirami, A., Geramifard, A., \& Kottur, S.
  (2021).
\newblock Dvd: A diagnostic dataset for multi-step reasoning in video grounded
  dialogue.
\newblock In {\em Proceedings of the 59th Annual Meeting of the Association for
  Computational Linguistics and the 11th International Joint Conference on
  Natural Language Processing (Volume 1: Long Papers)\/}, (pp. 5651--5665).

\bibitem[{Leidner \& Plachouras(2017)}]{leidner2017ethical}
Leidner, J.~L., \& Plachouras, V. (2017).
\newblock Ethical by design: Ethics best practices for natural language
  processing.
\newblock In {\em Proceedings of the First ACL Workshop on Ethics in Natural
  Language Processing\/}, (pp. 30--40).

\bibitem[{London(2021)}]{london2021common}
London, A.~J. (2021).
\newblock {\em For the Common Good: Philosophical Foundations of Research
  Ethics\/}.
\newblock Oxford University Press.

\bibitem[{London \& Kimmelman(2019)}]{london2019clinical}
London, A.~J., \& Kimmelman, J. (2019).
\newblock Clinical trial portfolios: a critical oversight in human research
  ethics, drug regulation, and policy.
\newblock {\em Hastings Center Report\/}, {\em 49\/}(4), 31--41.

\bibitem[{London et~al.(2020)London, Taljaard, \& Weijer}]{london2020loopholes}
London, A.~J., Taljaard, M., \& Weijer, C. (2020).
\newblock Loopholes in the research ethics system? informed consent waivers in
  cluster randomized trials with individual-level intervention.
\newblock {\em Ethics \& human research\/}, {\em 42\/}(6), 21--28.

\bibitem[{Mihaylov et~al.(2018)Mihaylov, Clark, Khot, \&
  Sabharwal}]{mihaylov2018can}
Mihaylov, T., Clark, P., Khot, T., \& Sabharwal, A. (2018).
\newblock Can a suit of armor conduct electricity? a new dataset for open book
  question answering.
\newblock In {\em Proceedings of the 2018 Conference on Empirical Methods in
  Natural Language Processing\/}, (pp. 2381--2391).

\bibitem[{Nekoto et~al.(2020)Nekoto, Marivate, Matsila, Fasubaa, Fagbohungbe,
  Akinola, Muhammad, Kabenamualu, Osei, Sackey
  et~al.}]{nekoto2020participatory}
Nekoto, W., Marivate, V., Matsila, T., Fasubaa, T., Fagbohungbe, T., Akinola,
  S.~O., Muhammad, S., Kabenamualu, S.~K., Osei, S., Sackey, F., et~al. (2020).
\newblock Participatory research for low-resourced machine translation: A case
  study in african languages.
\newblock In {\em Proceedings of the 2020 Conference on Empirical Methods in
  Natural Language Processing: Findings\/}, (pp. 2144--2160).

\bibitem[{Ning et~al.(2020)Ning, Wu, Dasigi, Dua, Gardner, Logan~IV, Marasovic,
  \& Nie}]{ning2020easy}
Ning, Q., Wu, H., Dasigi, P., Dua, D., Gardner, M., Logan~IV, R.~L., Marasovic,
  A., \& Nie, Z. (2020).
\newblock Easy, reproducible and quality-controlled data collection with
  crowdaq.
\newblock {\em arXiv preprint arXiv:2010.06694\/}.

\bibitem[{Orife et~al.(2020)Orife, Kreutzer, Sibanda, Whitenack, Siminyu,
  Martinus, Ali, Abbott, Marivate, Kabongo et~al.}]{orife2020masakhane}
Orife, I., Kreutzer, J., Sibanda, B., Whitenack, D., Siminyu, K., Martinus, L.,
  Ali, J.~T., Abbott, J., Marivate, V., Kabongo, S., et~al. (2020).
\newblock Masakhane--machine translation for africa.
\newblock {\em arXiv preprint arXiv:2003.11529\/}.

\bibitem[{P{\'e}rez-Rosas \& Mihalcea(2015)}]{perez2015experiments}
P{\'e}rez-Rosas, V., \& Mihalcea, R. (2015).
\newblock Experiments in open domain deception detection.
\newblock In {\em Proceedings of the 2015 Conference on Empirical Methods in
  Natural Language Processing\/}, (pp. 1120--1125).

\bibitem[{Rudinger et~al.(2018)Rudinger, Naradowsky, Leonard, \&
  Van~Durme}]{rudinger2018gender}
Rudinger, R., Naradowsky, J., Leonard, B., \& Van~Durme, B. (2018).
\newblock Gender bias in coreference resolution.
\newblock In {\em Proceedings of the 2018 Conference of the North American
  Chapter of the Association for Computational Linguistics: Human Language
  Technologies, Volume 2 (Short Papers)\/}, (pp. 8--14).

\bibitem[{Sheng \& Zhang(2019)}]{sheng2019machine}
Sheng, V.~S., \& Zhang, J. (2019).
\newblock Machine learning with crowdsourcing: A brief summary of the past
  research and future directions.
\newblock In {\em Proceedings of the AAAI Conference on Artificial
  Intelligence\/}, vol.~33, (pp. 9837--9843).

\bibitem[{Shmueli et~al.(2021)Shmueli, Fell, Ray, \& Ku}]{shmueli2021beyond}
Shmueli, B., Fell, J., Ray, S., \& Ku, L.-W. (2021).
\newblock Beyond fair pay: Ethical implications of nlp crowdsourcing.
\newblock In {\em Proceedings of the 2021 Conference of the North American
  Chapter of the Association for Computational Linguistics: Human Language
  Technologies\/}, (pp. 3758--3769).

\bibitem[{Silberman et~al.(2018)Silberman, Tomlinson, LaPlante, Ross, Irani, \&
  Zaldivar}]{silberman2018responsible}
Silberman, M.~S., Tomlinson, B., LaPlante, R., Ross, J., Irani, L., \&
  Zaldivar, A. (2018).
\newblock Responsible research with crowds: pay crowdworkers at least minimum
  wage.
\newblock {\em Communications of the ACM\/}, {\em 61\/}(3), 39--41.

\bibitem[{Strubell et~al.(2020)Strubell, Ganesh, \&
  McCallum}]{strubell2020energy}
Strubell, E., Ganesh, A., \& McCallum, A. (2020).
\newblock Energy and policy considerations for modern deep learning research.
\newblock In {\em Proceedings of the AAAI Conference on Artificial
  Intelligence\/}, vol.~34, (pp. 13693--13696).

\bibitem[{Taboada et~al.(2011)Taboada, Brooke, Tofiloski, Voll, \&
  Stede}]{taboada2011lexicon}
Taboada, M., Brooke, J., Tofiloski, M., Voll, K., \& Stede, M. (2011).
\newblock Lexicon-based methods for sentiment analysis.
\newblock {\em Computational linguistics\/}, {\em 37\/}(2), 267--307.

\bibitem[{Talmor et~al.(2019)Talmor, Herzig, Lourie, \&
  Berant}]{talmor2019commonsenseqa}
Talmor, A., Herzig, J., Lourie, N., \& Berant, J. (2019).
\newblock Commonsenseqa: A question answering challenge targeting commonsense
  knowledge.
\newblock In {\em Proceedings of the 2019 Conference of the North American
  Chapter of the Association for Computational Linguistics: Human Language
  Technologies, Volume 1 (Long and Short Papers)\/}, (pp. 4149--4158).

\bibitem[{Vaughan(2017)}]{vaughan2017making}
Vaughan, J.~W. (2017).
\newblock Making better use of the crowd: How crowdsourcing can advance machine
  learning research.
\newblock {\em J. Mach. Learn. Res.\/}, {\em 18\/}(1), 7026--7071.

\bibitem[{Whiting et~al.(2019)Whiting, Hugh, \& Bernstein}]{whiting2019fair}
Whiting, M.~E., Hugh, G., \& Bernstein, M.~S. (2019).
\newblock Fair work: Crowd work minimum wage with one line of code.
\newblock In {\em Proceedings of the AAAI Conference on Human Computation and
  Crowdsourcing\/}, vol.~7, (pp. 197--206).

\bibitem[{Williams et~al.(2018)Williams, Nangia, \& Bowman}]{williams2018broad}
Williams, A., Nangia, N., \& Bowman, S. (2018).
\newblock A broad-coverage challenge corpus for sentence understanding through
  inference.
\newblock In {\em Proceedings of the 2018 Conference of the North American
  Chapter of the Association for Computational Linguistics: Human Language
  Technologies, Volume 1 (Long Papers)\/}, (pp. 1112--1122).

\bibitem[{Xu et~al.(2020)Xu, Ju, Li, Boureau, Weston, \& Dinan}]{xu2020recipes}
Xu, J., Ju, D., Li, M., Boureau, Y.-L., Weston, J., \& Dinan, E. (2020).
\newblock Recipes for safety in open-domain chatbots.
\newblock {\em arXiv preprint arXiv:2010.07079\/}.

\bibitem[{Zaidan et~al.(2007)Zaidan, Eisner, \& Piatko}]{zaidan2007using}
Zaidan, O., Eisner, J., \& Piatko, C. (2007).
\newblock Using {``}annotator rationales{''} to improve machine learning for
  text categorization.
\newblock In {\em Human Language Technologies 2007: The Conference of the North
  {A}merican Chapter of the Association for Computational Linguistics;
  Proceedings of the Main Conference\/}, (pp. 260--267). Association for
  Computational Linguistics.
\newline\urlprefix\url{https://aclanthology.org/N07-1033}

\bibitem[{Zang et~al.(2021)Zang, Liu, Wang, Song, Zhang, \&
  Chen}]{zang2021photochat}
Zang, X., Liu, L., Wang, M., Song, Y., Zhang, H., \& Chen, J. (2021).
\newblock Photochat: A human-human dialogue dataset with photo sharing behavior
  for joint image-text modeling.
\newblock In {\em Proceedings of the 59th Annual Meeting of the Association for
  Computational Linguistics and the 11th International Joint Conference on
  Natural Language Processing (Volume 1: Long Papers)\/}, (pp. 6142--6152).

\bibitem[{Zhang et~al.(2021)Zhang, Zhang, Zhang, Yang, \&
  Lin}]{zhang2021multimet}
Zhang, D., Zhang, M., Zhang, H., Yang, L., \& Lin, H. (2021).
\newblock Multimet: A multimodal dataset for metaphor understanding.
\newblock In {\em Proceedings of the 59th Annual Meeting of the Association for
  Computational Linguistics and the 11th International Joint Conference on
  Natural Language Processing (Volume 1: Long Papers)\/}, (pp. 3214--3225).

\end{thebibliography}

\end{document}